\documentclass[twocolumn,showpacs,showkeys,preprintnumbers]{revtex4-1}
\usepackage{amssymb}

\usepackage{amsfonts}

\usepackage{latexsym}

\usepackage{dcolumn}

\usepackage{amsmath}

\usepackage{graphicx}

\usepackage{psfrag,pstricks}
\begin{document}

\title{Controllability of optical bistability, cooling and entanglement in hybrid cavity optomechanical systems by nonlinear atom-atom interaction}

\author{A. Dalafi$^{1}$ }
\email{adalafi@yahoo.co.uk}

\author{M. H. Naderi$^{1,2}$}
\author{ M. Soltanolkotabi$^{1,2}$}

\affiliation{$^{1}$ Department of Physics, Faculty of Science, University of Isfahan, Hezar Jerib, 81746-73441, Isfahan, Iran\\
$^{2}$Quantum Optics Group, Department of Physics, Faculty of Science, University of Isfahan, Hezar Jerib, 81746-73441, Isfahan, Iran}

\date{\today}

\begin{abstract}

We investigate the effects of atomic collisions as well as optomechanical mirror-field coupling on the optical bistability in a hybrid system consisting of a Bose-Einstein condensate inside a driven optical cavity with a moving end mirror. It is shown that the bistability of the system can be controlled by the \textit{s}-wave scattering frequency which can provide the possibility of realizing a controllable optical switch. On the other hand, by studying the effect of the Bogoliubov mode, as a secondary mechanical mode relative to the mirror vibrations, on the cooling process as well as the bipartite mirror-field and atom-field entanglements we find an interpretation for the cooling of the Bogoliubov mode. The advantage of this hybrid system in comparison to the bare optomecanical cavity with a two-mode moving mirror is the controllability of the frequency of the secondary mode through the \textit{s}-wave scattering interaction.

\end{abstract}


\maketitle

\section{Introduction}
%
% 1st draft
%
In recent years, optomechanical cooling of a mechanical resonator close to its quantum mechanical ground state has been an interesting topic for a wide range of fields of physics such as ultrahigh precision measurements \cite{LaHaye} and the detection of gravitational waves \cite{Abramovici}. It has also provided a good approach for fundamental studies of the transition between the quantum and the classical world \cite{marshall}. This phenomenon can be realized in the optomechanical systems consisting of  an optical cavity with a movable end mirror or with a membrane in the middle \cite{Genes 2008, Teufel, Liberato, Barzanjeh2}.

On the other hand an effective optomechanical coupling can be simulated by a Bose-Einstein condensate (BEC) trapped inside a high finesse optical cavity \cite{Brenn Nature,Gupta,Brenn Science}. In the dispersive regime where the laser pump is far detuned from the atomic resonance the excited electronic state of the atoms can be adiabatically eliminated and consequently the only degrees of freedom of atoms will be their mechanical motions \cite{Masch Ritch 2004,Dom JB}. For low photon numbers or in the weakly interacting regime the dynamics can be restricted to the first motional mode which plays the role of the mechanical oscillator\cite{Kanamoto 2010, Nagy Ritsch 2009}.  

In this paper we study	 a hybrid optomechanical system consisting of a BEC trapped inside an optical cavity with a moving end mirror which is driven through its fixed mirror. It is shown that in the weak coupling regime and under the Bogoliubov approximation the side mode of the BEC is coupled to the optical field through a radiation pressure term just like what occurs for the vibrational mode of the mirror. Besides, we assume that the BEC has been isolated from its environment so that its effective temperature is several orders of magnitude smaller than the equilibrium temperature of the mirror. This system is equivalent to a bare optomechanical system with a moving end mirror whose  two vibrational modes are coupled with the radiation pressure of the cavity. It has been shown \cite{Genes2008} that in such a system the cooling of the main mode depends on the frequency difference between the two modes. When this difference is equal or larger than the main mode frequency, the secondary mode does not disturb the cooling of the main mode. In the hybrid system considered here, the Bogoliubov mode of the BEC plays the role of the secondary vibrational mode in the bare optomechanical one.

One of the advantages of the hybrid system with respect to the bare one is that the frequency of the secondary mode (Bogoliubov mode) depends on the \textit{s}-wave scattering frequency $\omega_{sw}$ of the nonlinear atom-atom interaction which is controllable through the transverse trapping frequency $\omega_{\mathrm{\perp}}$ \cite{Morsch}. In this way the frequency difference $\Delta\omega$ between the two modes can be controlled. Furthermore, in the present setup we can introduce an interpretation for the cooling process of the Bogoliubov mode through its correspondence with the mirror vibrational mode and the way it gets entangled with the optical field.

On the other hand the nonlinear coupling between the two modes and the optical field (radiation pressure coupling) results in the optical bistability of the system where there are two stable mean-field solutions corresponding to different photon numbers and oscillator displacements \cite{meys}. For an optical cavity containing a BEC with no moving mirror this phenomenon is observable even below the single photon level \cite{ Ritter Appl. Phys. B, Szirmai 2010}. Here, we also investigate both the effects of \textit{s}-wave scattering interaction and the mirror-field coupling on the optical bistability of the system.

Recently the optical bistability behavior has been investigated for a system consisting of a BEC trapped inside a Fabry-Perot cavity with fixed mirrors which is pumped both along the cavity axis and from the transverse side \cite{Zubairy}. It has been shown that the optical bistability can be controlled by this transverse pumping field. For small values of transverse pumping, there exists a clear bistability for a particular range of the input pump along the cavity axis while increasing the transverse pumping field, the range of the bistability region is reduced. Increasing the transverse pumping above a critical value causes the bistable behavior to be disappeared. It has been shown that this phenomenon can be exploited for realization an optical switch in which the transverse pump is used as a control parameter to enable or disable the switch.

In this work, we introduce another possibility of an optical switch realization in an optomechanical cavity with no necessity of using a transverse pumping field. It is shown that the bistability of the system can be controlled  by the \textit{s}-wave scattering frequency $\omega_{sw}$ which itself is controllable by the transverse trapping frequency $\omega_{\mathrm{\perp}}$. Here, the optical bistability of the system does not disappear but it is shifted toward the larger values of the parallel pumping field. 
  
The paper is structured as follows. In Sec. II we derive the Hamiltonian of the system consisting of a BEC inside an optomechanical cavity with a moving end mirror in the discrete mode approximation considering \textit{s}-wave scattering interaction. In Sec. III the quantum Langevin equations (QLEs) are derived and linearized around the semiclassical steady state. In Sec. IV we study the mean-field solutions and investigate the effects of \textit{s}-wave scattering interaction and the the mirror-field coupling on the optical bistability of the system. In Sec. V we examine the effect of the Bogoliubov mode as the secondary mode on the cooling and entanglement of the vibrational mode of the mirror (main mode). Finally, our conclusions are summarized in Sec. VI.
%----------SECTION----------%
\section{System Hamiltonian}

\begin{figure}[ht]
\centering
\includegraphics[scale=1]{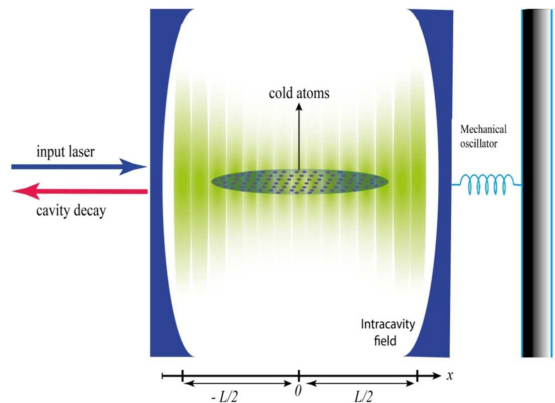} 
\caption{(Color online) A BEC trapped in an optomechanical cavity interacting dispersively with a single cavity mode. The cavity which decays at rate $\kappa$ is driven through the fixed mirror by a laser with frequency $\omega_{p}$ and the end mirror is free to oscillate at mechanical frequency $\omega_{m}$.}
\label{fig:fig1}
\end{figure}

We consider a system consisting of a BEC of $N$ two-level atoms with mass $m_{0}$ and transition freqyency $\omega_{a}$ inside an optomechanical cavity with length $L$ whose end mirror is free to oscillate at mechanical frequency $\omega_{m}$. The cavity is driven at rate $\eta=\sqrt{2\mathcal{P}\kappa/\hbar\omega_{c}}$ through the fixed mirror by a laser with frequency $\omega_{p}$, and wavenumber $k=\omega_{p}/c$ ($\mathcal{P}$ is the laser power and $\kappa$ is the cavity decay rate). We assume the BEC to be confined in a cylindrically symmetric trap with a transverse trapping frequency $\omega_{\mathrm{\perp}}$ and negligible longitudinal confinement along the $x$ direction (Fig.\ref{fig:fig1}). In this way we can describe the dynamics within an effective one-dimensional model by quantizing the atomic motional degree of freedom along the $x$ axis only.

In the dispersive regime where the laser pump is far detuned from the atomic resonance ($\Delta_{a}=\omega_{p}-\omega_{a}$  exceeds the atomic linewidth $\gamma$ by orders of magnitude), the excited electronic state of the atoms can be adiabatically eliminated and spontaneous emission can be neglected \cite{Masch Ritch 2004}. In the frame rotating at the pump frequency, the many-body Hamiltonian reads
\begin{eqnarray}\label{H1}
H&=&\int_{-L/2}^{L/2} dx \Psi^{\dagger}(x)\Big[\frac{-\hbar^{2}}{2m_{0}}\frac{d^{2}}{dx^{2}}+\hbar U_{0} \cos^2(kx) a^{\dagger} a\nonumber\\
&&+\frac{1}{2} U_{s}\Psi^{\dagger}(x)\Psi(x)\Big] \Psi(x)+\hbar\Delta_{c} a^{\dagger} a + i\hbar\eta  (a^{\dagger}-a)\nonumber\\
&&+\frac{1}{2}\hbar\omega_{m}(p^{2}+q^{2})-\hbar\xi a^{\dagger}a q,
\end{eqnarray}
where $a$ is the annihilation operator for a cavity photon, $\Delta_{c}=\omega_{c}-\omega_{p}$ is the cavity-pump detuning, $U_{0}=g_{0}^{2}/\Delta_{a}$ is the optical lattice barrier height per photon which represents the atomic backaction on the field, $g_{0}$ is the vacuum Rabi frequency, $U_{s}=\frac{4\pi\hbar^{2} a_{s}}{m_{0}}$ and $a_{s}$ is the two-body \textit{s}-wave scattering length \cite{Masch Ritch 2004,Dom JB}. The last two terms in the Hamiltonian of Eq.(\ref{H1}) represent, respectively, the energy of the mechanical mode and the radiation pressure coupling of rate $\xi=(\omega_{c}/L)\sqrt{\hbar/m\omega_{m}}$ where $m$ is the effective mass of the moving mirror.

In the weakly interacting regime, only the first two symmetric momentum side modes with momenta $\pm2\hbar k$ are excited by fluctuations resulting from the atom-light interaction \cite{Nagy Ritsch 2009}.  In this way because of the parity conservation and considering the Bogoliubov approximation, the atomic field operator can be expanded as
\begin{equation}\label{opaf}
\Psi(x)=\sqrt{\frac{N}{L}}+\sqrt{\frac{2}{L}}\cos(2kx) c.
\end{equation}

In the case that the system does not have parity symmetry, for example, when the BEC is inside a ring cavity, one should also consider sine modes which in our model have been set aside \cite{Steinke Meustre 2011,Chen Meystre 2010}. By substituting the atomic field operator, Eq.(\ref{opaf}), into the Hamiltonian of Eq.(\ref{H1}) and introducing the Bogoliubov mode quadratures $Q=(c+c^{\dagger})/\sqrt{2}$ and $P=(c-c^{\dagger})/\sqrt{2}i$ one can arrive at the following Hamiltonian:
\begin{eqnarray}\label{OpA}
H&=&\hbar \delta_{c} a^{\dagger}a+i\hbar\eta (a^{\dagger}-a)+\frac{1}{2}\hbar\omega_{m}(p^{2}+q^{2})-\hbar\xi a^{\dagger}a q\nonumber\\
&&+\frac{1}{2}\hbar\Omega_{c}(P^2+Q^2)+\hbar\zeta a^{\dagger}a Q+\frac{1}{2}\hbar\omega_{sw} Q^2,
\end{eqnarray}
where $\delta_{c}=\Delta_{c}+\frac{1}{2}N U_{0}$ is the effective Stark-shifted detuning, $\Omega_{c}=4\omega_{R}+\frac{1}{2}\omega_{sw}$ ($\omega_{R}$, the recoil frequency of the condensate atoms) and $\zeta=\frac{1}{2}\sqrt{N}U_{0}$ is the strength of interaction between the BEC mode and the intracavity field. 
The first two terms in the second line of Eq.(\ref{OpA}) denote the energy of the Bogoliubov mode and the radiation pressure coupling of the Bogoliubov mode and the optical field, respectively. The last term is the atom-atom interaction energy where $\omega_{sw}=8\pi\hbar a_{s}N/m_{0}Lw^2$  is the \textit{s}-wave scattering frequency and $w$ is the waist of the optical potential.

\section{Dyanamics of The System}
The degrees of freedom of the system can be represented by the vector $u=[X, Y, q, p, Q, P,]^{T}$ where $X=(a+a^{\dagger})/\sqrt{2}$ and $Y=(a-a^{\dagger})/\sqrt{2}i$ are the optical field quadratures. Under intense laser pumping these operators can be linearized and expanded around their respective classical mean values $u_{s,j}$ as $u_{j}=u_{s,j}+\delta u_{j}$ where $\delta u_{j}$ are zero-mean fluctuation operators. The mean values can be determined by solving the steady-state Langevin equations, which will lead to the following equations:
\begin{subequations}\label{ss}
\begin{eqnarray}
\alpha&=&\frac{\eta}{\sqrt{\Delta^2+\kappa^2}},\\
Q_{s}&=&-\frac{\zeta\alpha^2}{\Omega_{c}+\omega_{sw}+\frac{\gamma_{c}^2}{\Omega_{c}}},\\
P_{s}&=&\frac{\gamma_{c}}{\Omega_{c}}Q_{s},\\
q_{s}&=&\frac{\xi}{\omega_{m}}\alpha^2,  p_{s}=0,
\end{eqnarray}
\end{subequations}
where we have assumed the optical field mean value, $\alpha$, to be a real number \cite{Konya Domokos 2011} and $\gamma_{c}$ is the dissipation of the collective density excitations of the BEC and $\Delta=\delta_{c}-\xi q_{s}+\zeta Q_{s}$ is the effective detuning.  

The dynamics of the quantum fluctuations can be described by the linearized QLEs which can be written in the compact matrix form,
\begin{equation}\label{nA}
\delta\dot{u}(t)=A \delta u(t)+n(t),
\end{equation}
where $\delta u=[\delta X,\delta Y, \delta q,\delta p,\delta Q,\delta P]^{T}$ is the vector of continuous variable fluctuation operators and
\begin{equation}
n(t)=[X_{in}(t),Y_{in}(t),0,f_{m}(t),f_{Q}(t),f_{P}(t)]^{T},
\end{equation}
is the corresponding vector of noises. The cavity-field quantum vacuum fluctuation is accounted for by the input-noise operators $X_{in}=(a_{in}+a_{in}^{\dagger})/\sqrt{2}$ and $Y_{in}=(a_{in}-a_{in}^{\dagger})/\sqrt{2}i$ where $a_{in}(t)$ satisfies the Markovian correlation functions, i.e., $\langle a_{in}(t)a_{in}^{\dagger}(t^{\prime})\rangle=(n_{ph}+1)\delta(t-t^{\prime})$, $\langle a_{in}^{\dagger}(t^{\prime})a_{in}(t)\rangle=n_{ph}\delta(t-t^{\prime})$ with the average thermal photon number $n_{ph}$ which is nearly zero at optical frequencies \cite{Gardiner}. Besides, the Brownian noise operator, i.e., $f_{m}$, accounts for the mechanical noise that couples into the mode of the mirror vibration from the thermal environment at temperature $T$. In the limit of a high mechanical quality factor it can be considered as a Markovian noise. On the other hand, $f_{Q}(t)$ and $f_{P}(t)$ are the thermal noise inputs for the side mode of BEC which also satisfy the same Markovian correlation functions as those of the optical noise. Here, we have assumed that the BEC has been trapped and isolated from its environment so that its effective temperature $T_{c}$ is several orders of magnitude smaller than the equilibrium temperature $T$ of the mirror.  The noise sources are assumed uncorrelated for the different modes of both the matter and light fields. 
 The $6\times6$ matrix $A$ is the drift matrix given by
\begin{equation}
A=\left(\begin{array}{cccccc}
-\kappa & \Delta & 0 & 0 & 0 & 0 \\
   -\Delta & -\kappa &\sqrt{2}\xi\alpha &0 &-\sqrt{2}\zeta\alpha & 0 \\
    0 & 0 & 0 & \omega_{m} & 0& 0 \\
  \sqrt{2}\xi\alpha & 0 & -\omega_{m} & -\gamma_{m} & 0&0\\
  0& 0& 0 &0 &-\gamma_{c} & \Omega_{c} \\
   -\sqrt{2}\zeta\alpha & 0 & 0 & 0 & -(\Omega_{c}+\omega_{sw}) &-\gamma_{c}
  \end{array}\right).
\label{A}
\end{equation}

The system is stable only if the real part of all the eigenvalues of the matrix $A$ are negative. These stability conditions can be obtained by using the Routh-Hurwitz criterion \cite{RH}. Due to the linearized dynamics of the fluctuations and since all noises are Gaussian the steady state is a zero-mean Gaussian state which is fully characterized by the $6\times6$ stationary correlation matrix (CM) $V$, with components $V_{ij}=\langle \delta u_i(\infty)\delta u_j(\infty)+\delta u_j(\infty)\delta u_i(\infty)\rangle/2 $. Using the Langevin equations, one can show that V fulfills the  Lyapunov equation \cite{Genes2008}
\begin{equation}\label{lyap}
AV+VA^T=-D,
\end{equation}
where
\begin{equation}\label{D}
 D=\mathrm{Diag}[\kappa,\kappa,0,\gamma_{m}(2n+1),\gamma_{c},\gamma_{c}]
\end{equation}
is the diffusion matrix, with $n=[\mathrm{exp}(\hbar \omega_m/k_B T)-1]^{-1}$ as the mean number of thermal excitations of the mechanical mode. Equation(\ref{lyap}) is linear in $V$ and can be straightforwardly solved. However, the explicit form of $ V $ is complicated and is not reported here.

\section{The mean-field solution and Optical Bistability}

In this section we discuss our results based on the numerical solutions of Eq.(\ref{ss}) for the mean fields and we will show how the physical parameters of the system like the atom-atom interaction strength and the optomechanical coupling ($\xi$) affect the optical bistability of the system. 

We analyse our results based on the experimentally feasible parameters given in Refs.\cite{Gigan,Arcizet,Rgenes}, i.e., an optomechanical cavity of length $L$=1 mm with bare frequency $\omega_{c}$ corresponding to a wavelength of $\lambda$=1064 nm with finesse $\mathcal{F}=3\times 10^4$ and the damping rate $\kappa=\pi c/L\mathcal{F}$. The end mirror with mass  $m=50$ ng and	 quality factor $\mathcal{Q}=10^5$ oscillates with frequency $\omega_{m}=2\pi\times10^7$ Hz. The optical mode is coherently driven at rate $\eta=\sqrt{2\mathcal{P}\kappa/\hbar\omega_{c}}$ and the recoil frequency of atoms is $\omega_{R}=0.1\omega_{m}$.  

\begin{figure}[ht]
\centering
\includegraphics[width=2.2in]{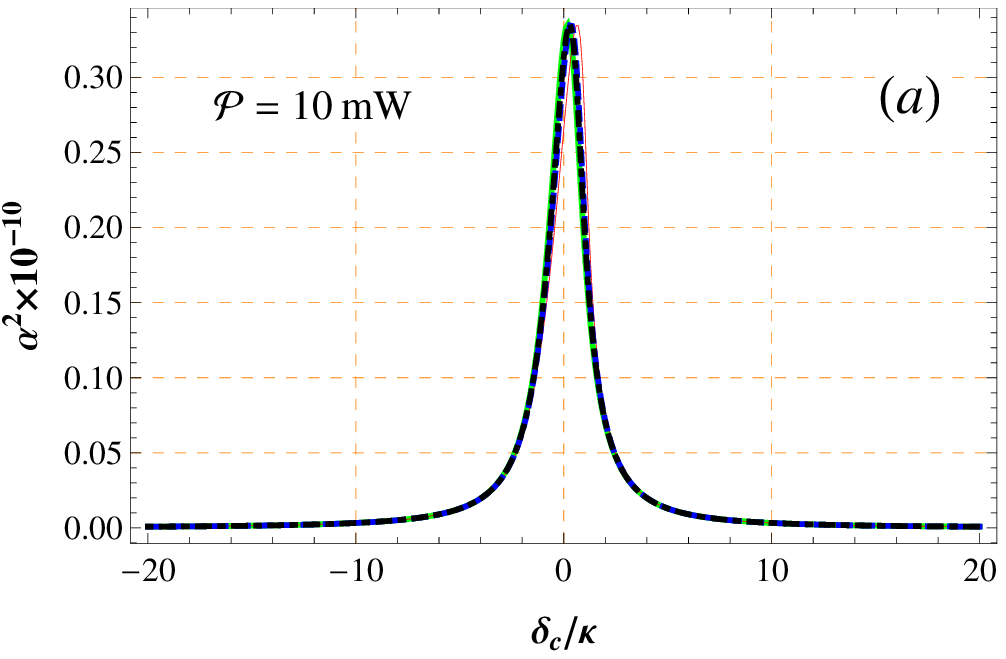}
\includegraphics[width=2.2in]{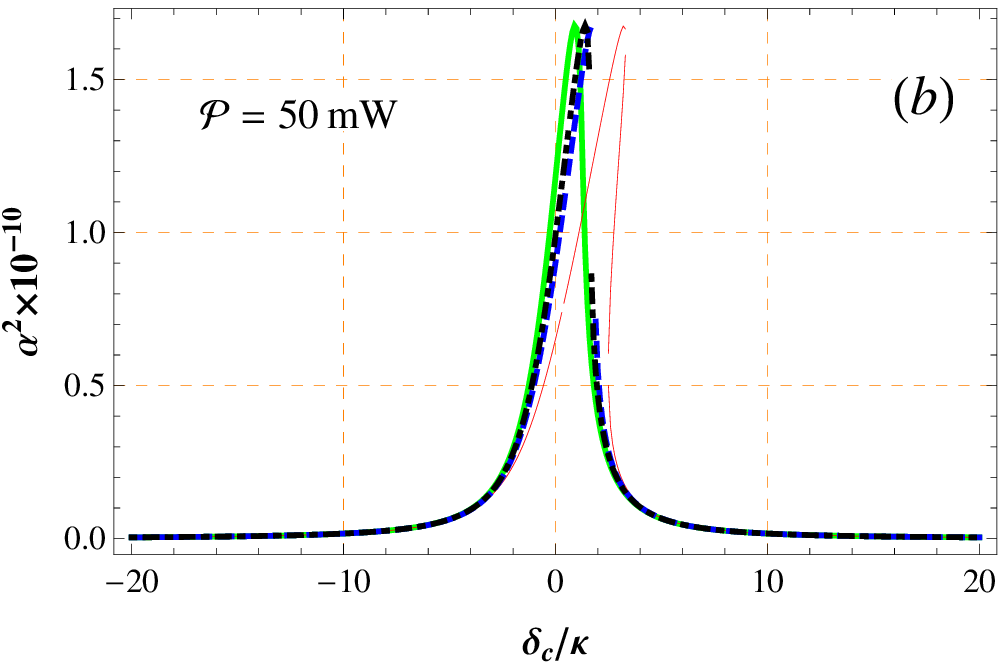}
\includegraphics[width=2.2in]{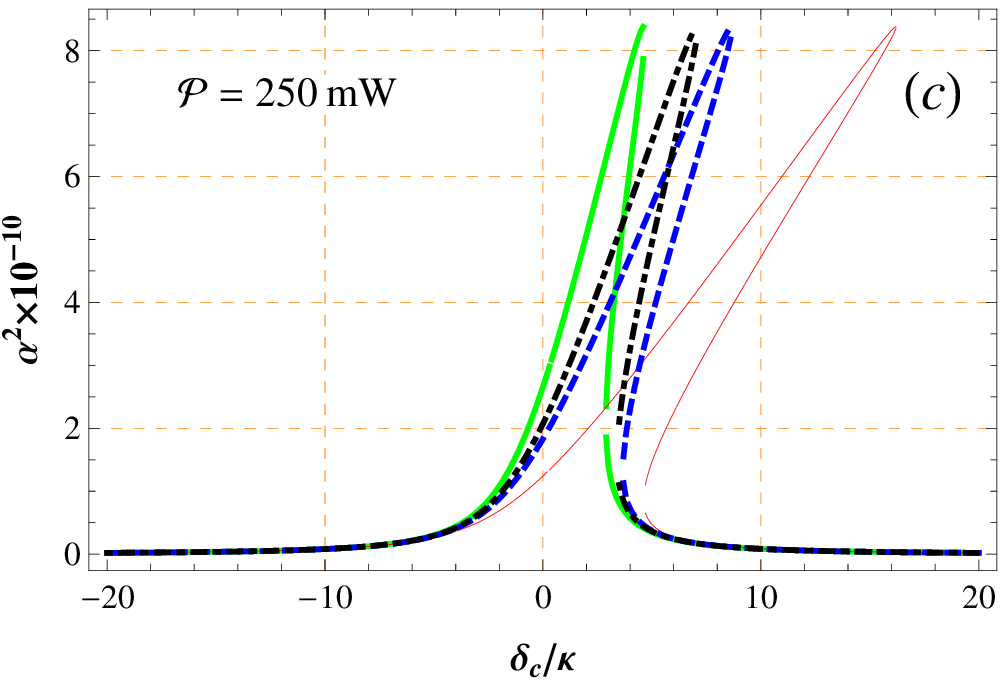}
\includegraphics[width=2.2in]{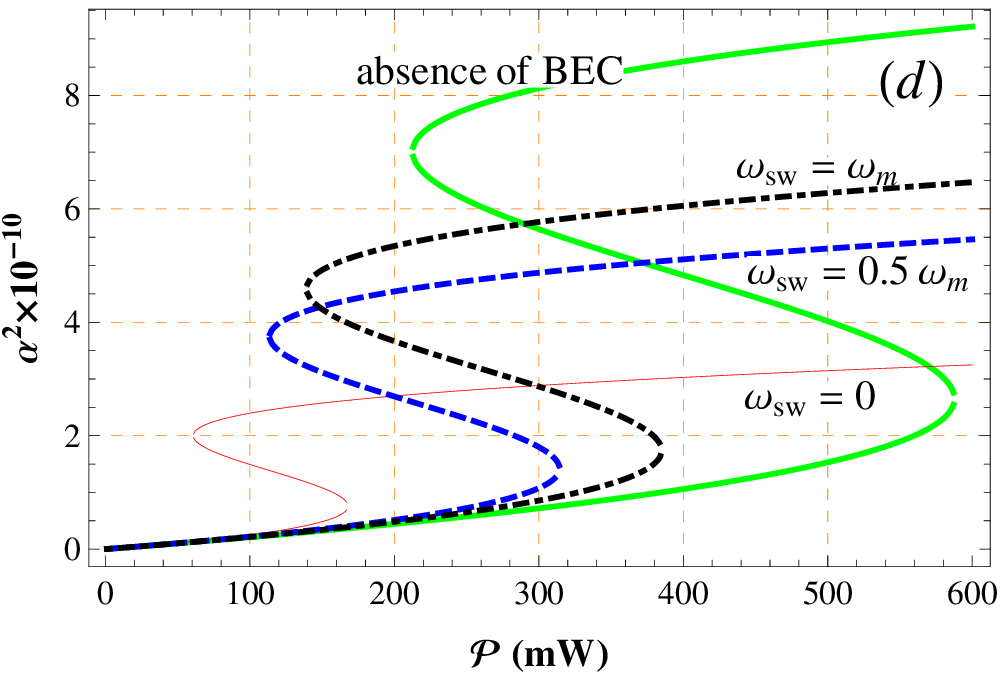}
\caption{
(Color online) Mean intracavity photon number $\alpha^2$ versus the normalized cavity-pump detuning $\delta_{c}/\kappa$ in the absence of BEC (green thick line) and in the presence of BEC for three different values of atomic interaction with $\omega_{sw}=0$ (red thin line), $\omega_{sw}=0.5 \omega_{m}$ (blue dashed line) and $\omega_{sw}=\omega_{m}$ (black dashed-dotted line), calculated for three different pump strengths: (a) $\mathcal{P}$=10 mW, (b) $\mathcal{P}$=50 mW, and (c) $\mathcal{P}$=250 mW. (d) Mean  photon number $\alpha^2$ versus pump strength at $\delta_{c}=4\kappa$. The cavity has a length of $L$=1 mm, a wavelength of $\lambda$=1064 nm with finesse $\mathcal{F}=3\times 10^4$ and the damping rate $\kappa=\pi c/L\mathcal{F}$. The end mirror of the cavity with mass $m=50$ ng oscillates with the natural frequency $\omega_{m}=2\pi\times10^7$ Hz.}
\label{fig:fig2}
\end{figure}

In Fig.\ref{fig:fig2} we have plotted the mean number of photons, $\alpha^2$, versus the normalized cavity-pump detuning $\delta_{c}/\kappa$ for three different pump strengths $\mathcal{P}$=10 mW (Fig.\ref{fig:fig2}a),  $\mathcal{P}$=50 mW (Fig.\ref{fig:fig2}b) and  $\mathcal{P}$=250 mW (Fig.\ref{fig:fig2}c). In order to see the effect of  BEC on the optical bistability of the system we have plotted the mean photon number of the system both in the absence of the BEC (green thick line) and in the presence of BEC with three different atom-atom interaction values: $\omega_{sw}=0$ (red thin line), $\omega_{sw}=0.5 \omega_{m}$ (blue dashed line) and $\omega_{sw}= \omega_{m}$ (black dashed-dotted line).

In Fig.\ref{fig:fig2} (a) the power of laser pump is $\mathcal{P}$=10 mW which is far below the bistability threshold of the system. So all four curves have been approximately overlapped. As is seen, neither the absence of BEC nor the atom-atom interaction strength in the presence of BEC affect the mean number of photons considerably. This result is similar to the one obtained in the Fig.2a of Ref.\cite{Dalafi} where a BEC has been considered inside a Fabry-Perot cavity with fixed mirrors and very weak pumping strength. It was shown that increasing the \textit{s}-wave scattering interaction shifts the resonance frequency of the cavity to the lower values (Similar results have been obtained from numerical solution of the Gross-Pitaevskii equation (GPE) \cite{Zhang 2009,Horak 2000}). Similarly here, the curves with non zero atom-atom interaction (dashed line and dashed-dotted line) have been shifted to the left in comparison to the red line.

\begin{figure}[ht]
\centering
\includegraphics[width=2.5in]{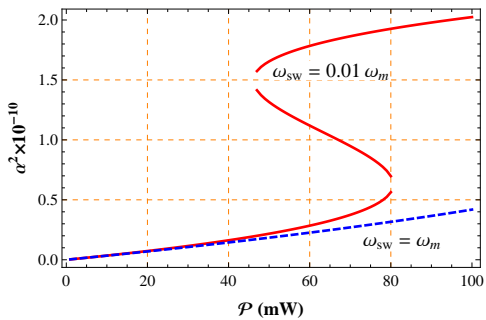}
\caption{
(Color online) Mean  photon number $\alpha^2$ versus the pump strength at $\delta_{c}=3\kappa$ for two different values of atom-atom interaction strength: $\omega_{sw}=\omega/100$ (red solid line) and $\omega_{sw}=\omega_{m}$ (blue dashed line). }
\label{fig:fig3}
\end{figure}

Increasing the pump strength causes the system to approach the bistability region and makes the effect of atom-atom interaction more observable. In Fig.\ref{fig:fig2} (b) where the power of the laser is $\mathcal{P}$=50 mW the system in the absence of BEC is still below the threshold (green thick line) while the system containing a non interacting BEC has just started to be bistable in a small region (red thin line). In contrast, increasing the atom-atom interaction has pushed the system to the stable region. In fact the \textit{s}-wave scattering interaction causes the bistability to happen at stronger pump powers. Here, the non interacting BEC (red thin line) is above the threshold while the interacting ones (dashed and dashed-dotted lines) are below the threshold. In Fig.\ref{fig:fig2} (c) where the pump strength has been increased to $\mathcal{P}$=250 mW the curves are completely resolved from each other and their bistability regions have become wider.

In order to examine the effect of \textit{s}-wave scattering interaction on optical bistability more clearly we have also plotted the mean number of intracavity photons versus the input laser power for a fixed value of $\delta_{c}=4\kappa$ in Fig.\ref{fig:fig2} (d). As is seen, the threshold of the system in the absence of BEC is slightly above 200 mW (green thick line) while it is about 60 mW for the system containing a non interacting BEC (red thin line) and about 110 mW and 140 mW for the system containing interacting BEC with $\omega_{sw}=0.5\omega_{m}$ (dashed line) and $\omega_{sw}= \omega_{m}$ (dashed-dotted line), respectively.

In short, the presence of a BEC inside an optomechanical cavity causes the system to become bistable at lower pump powers and also decreases the bistability region width (in terms of the parallel pump intensity). On the other hand, the atomic interaction acts so that to compensate this effect, i.e., pushes the threshold point to the upper values and increases the bistability region width. These results which have been obtained based on a discrete mode approximation (DMA) method, considering the Bogoliubov approximation and atom-atom interaction, are in good accordance with those obtained from numerical solution of the GPE (Fig. 6 of Ref.\cite{Zubairy}). Furthermore in a very recent work \cite{new nagy}, the effect of \textit{s}-wave scattering interaction on the optical bistability has been studied and it has been shown that the results obtained by considering the Bogoliubov approximation are satisfactory.

In Ref.\cite{Zubairy} the authors has shown the possibility of realizing a controllable optical switch by adding a perpendicular pump field to a Fabry-Perot cavity (with fixed mirrors and containing a BEC) to control the bistability of the cavity photon number.  They have discussed that the bistability of the cavity photon number with the change of the parallel pumping field can be controlled and suppressed by the perpendicular pumping field. In this way an optical switch can be realized by using the parallel pump field as the input and the perpendicular pump field as  the control.

\begin{figure}[ht]
\centering
\includegraphics[width=2.5in]{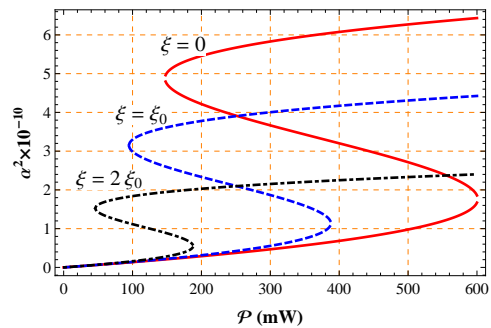}
\caption{ 
(Color online) Mean  photon number $\alpha^2$ versus the pump strength at $\delta_{c}=5\kappa$, $\zeta=\xi_{0}$ and $\omega_{sw}=0.1 \omega_{m}$ for three different values of the optomechanical coupling: $\xi=0$ (red solid line), $\xi=\xi_{0}$ (blue dashed line), and $\xi=2\xi_{0}$ (black dashed-dotted line) where $\xi_{0}=330$ Hz. }
\label{fig:fig4}
\end{figure}

Here we are going to introduce another possibility of an optical switch realization in an optomechanical cavity with no necessity of using a perpendicular pump field. Based on the results obtained in Fig.\ref{fig:fig2} (d) we can control the bistability of the system by the \textit{s}-wave scattering frequency ($\omega_{sw}$) which itself is controllable by the transverse trapping frequency $\omega_{\mathrm{\perp}}$ \cite{Morsch}. For clarity, we have represented this phenomenon separately in  Fig.\ref{fig:fig3} where the cavity mean photon number has been plotted versus the input laser power at $\delta_{c}=3\kappa$ for two values of $\omega_{sw}=0.01 \omega_{m}$ (red line) and $\omega_{sw}=\omega_{m}$ (blue dashed line). As is seen, if we restrict the pumping strength in the range 45-80 mW the system will be bistable for $\omega_{sw}=\omega_{m}/100$ while for $\omega_{sw}=\omega_{m}$ the system is not bistable (because the bistability region has been moved to the stronger powers). In this way we can enable or disable the optical switch using the transverse trapping frequency $\omega_{\mathrm{\perp}}$ .

In order to examine the effect of the optomechanical coupling parameter on the bistability behavior of the system we have plotted in Fig.\ref{fig:fig4} the mean photon number versus the input laser power for three different values of the optomechanical parameter $\xi$ for a system containing an interacting BEC with $\zeta=\xi_{0}$ and $\omega_{sw}=0.1\omega_{m}$ where $\xi_{0}$=330 Hz and the values of other parameters are the same as those in Fig.\ref{fig:fig1}. The red solid line belongs to a system with no optomechanical coupling (i.e., $m\rightarrow \infty$) while the other two lines correspond to $\xi=\xi_{0}$ (blue dashed line) and $\xi=2\xi_{0}$ (black dashed-dotted line). As is seen, increasing the optomechanical coupling causes the system to become bistable at lower pumping powers and also decreases the bistability region width  which is similar to the effect of the presence of a BEC in an optomechanical cavity but it acts in the reverse direction of the atom-atom interaction effect (see Fig.\ref{fig:fig2} (d)).

\section{Cooling and Entanglement}

After calculating the mean-field values we can obtain the elements of the drift matrix $A$ and solve for the steady sate solutions of Eq.(\ref{nA}). As explained before, by solving the Lyapunov equation [Eq.(\ref{lyap})] we can obtain the correlation matrix $V$ which gives us the second-order correlations of the fluctuations.  In this way we can calculate the effective incoherent number of mirror phonons,
\begin{equation}\label{deltanm}
\delta n_{m}=\frac{V_{33}+V_{44}-1}{2},
\end{equation}
and the effective incoherent number of atoms in the Bogoliubov mode,
\begin{equation}\label{deltanc}
\delta n_{c}=\frac{V_{55}+V_{66}-1}{2}.
\end{equation}
On the other hand the bipartite entanglement between the atomic and photonic degrees of freedom can also be calculated by using the logarithmic negativity \cite{eis}:
\begin{equation}\label{en}
E_N=\mathrm{max}[0,-\mathrm{ln} 2 \eta^-],
\end{equation}
where  $\eta^{-}\equiv2^{-1/2}\left[\Sigma(\mathcal{V}_{bp})-\sqrt{\Sigma(\mathcal{V}_{bp)}^2-4 \mathrm{det} \mathcal{V}_{bp}}\right]^{1/2}$  is the lowest symplectic eigenvalue of the partial transpose of the $4\times4$ CM, $\mathcal{V}_{bp}$, associated with the selected bipartition, obtained by neglecting the rows and columns of the uninteresting mode,
\begin{equation}\label{bp}
\mathcal{V}_{bp}=\left(
     \begin{array}{cc}
     \mathcal{B}&\mathcal{C}\\
      \mathcal{C}^{T}&\mathcal{B}^{\prime}\\
       \end{array}
   \right),
\end{equation}
and $\Sigma(\mathcal{V}_{bp})=\mathrm{det} \mathcal{B}+\mathrm{det} \mathcal{B}^{\prime}-2\mathrm{det} \mathcal{C}$.

\begin{figure}[ht]
\centering
\includegraphics[width=2.7in]{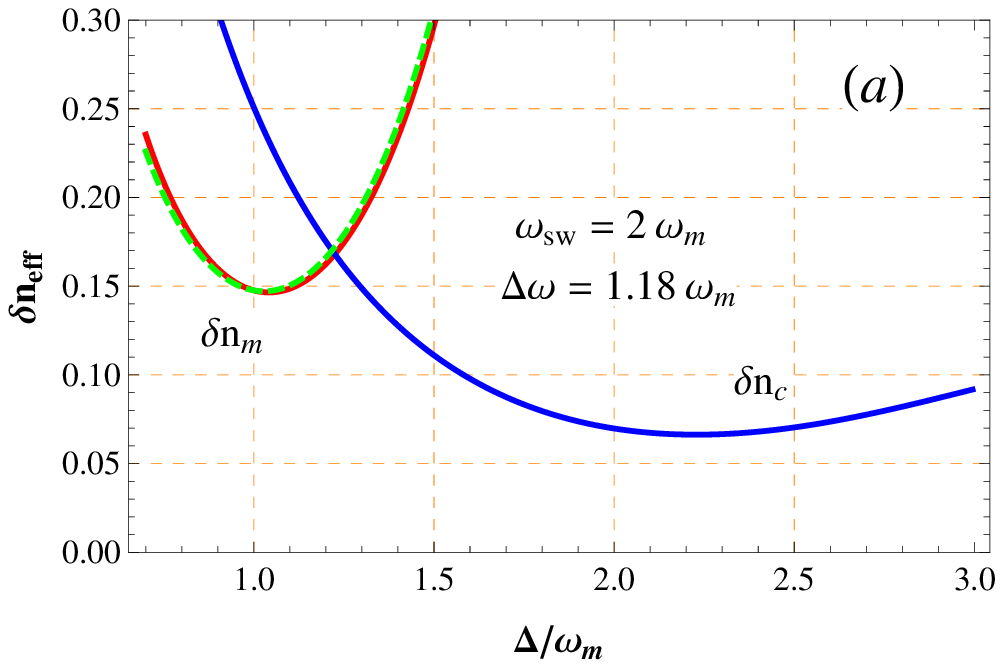}
\includegraphics[width=2.7in]{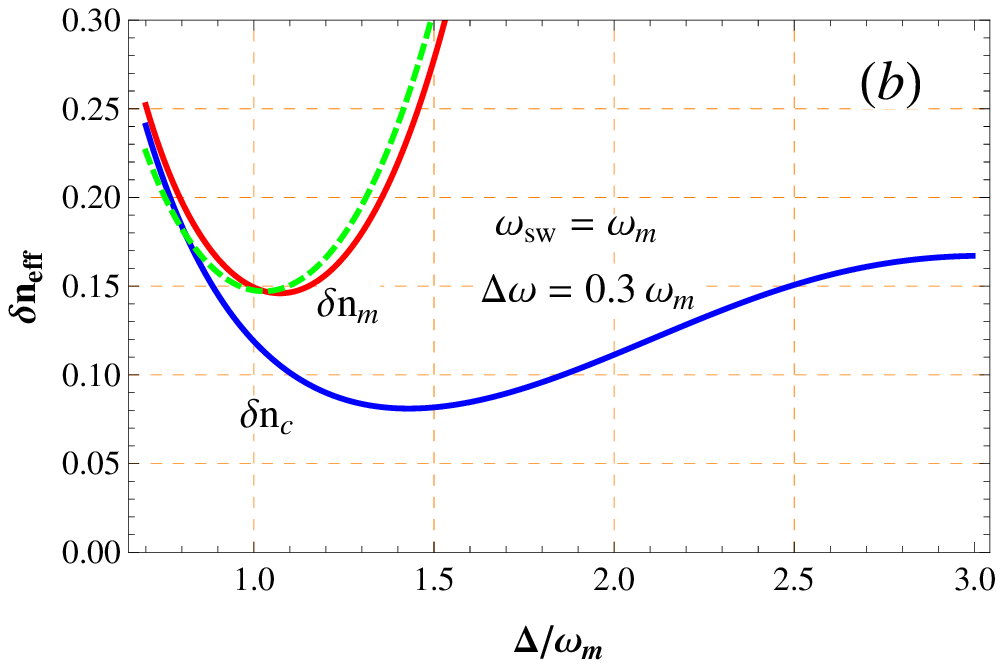}
\includegraphics[width=2.7in]{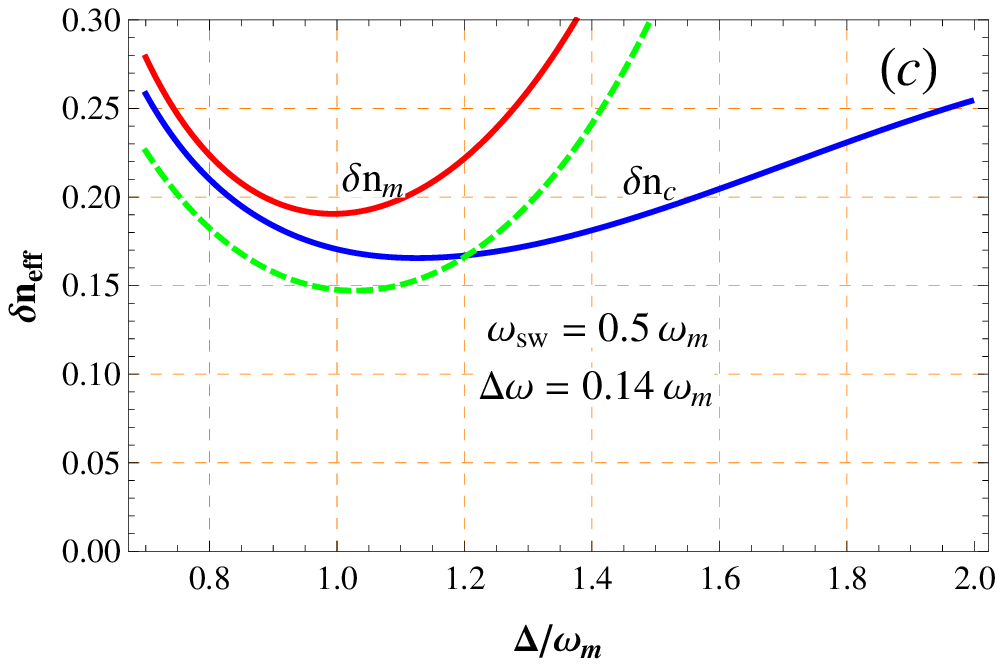}
\caption{
(Color online)
The effective incoherent number of mirror phonons $\delta n_{m}$ in the presence of BEC (red line) and in the absence of BEC (green dashed line) as well as the effective incoherent number of atoms in the Bogoliubov mode $\delta n_{c}$ (blue line) versus the normalized detuning $\Delta/\omega_{m}$ for three different values of atom-atom interaction strength: (a) $\omega_{sw}=2\omega_{m}$, (b) $\omega_{sw}=\omega_{m}$, and (c) $\omega_{sw}=0.5\omega_{m}$ which leads to three different frequency differences $\Delta\omega$ between the mechanical and the Bogoliubov modes. The effective temperature of the BEC is $T_{c}=0.1 \mu$K with the Bogoliubov mode damping rate $\gamma_{c}=0.001\kappa$, the initial reservoir temperature of the mirror is $T=0.4$K and $\mathcal{P}=50$ mW. The other parameters are the same as those of Fig.\ref{fig:fig2}.}
\label{fig:fig5}
\end{figure}

\begin{figure}[ht]
\centering
\includegraphics[width=2.7in]{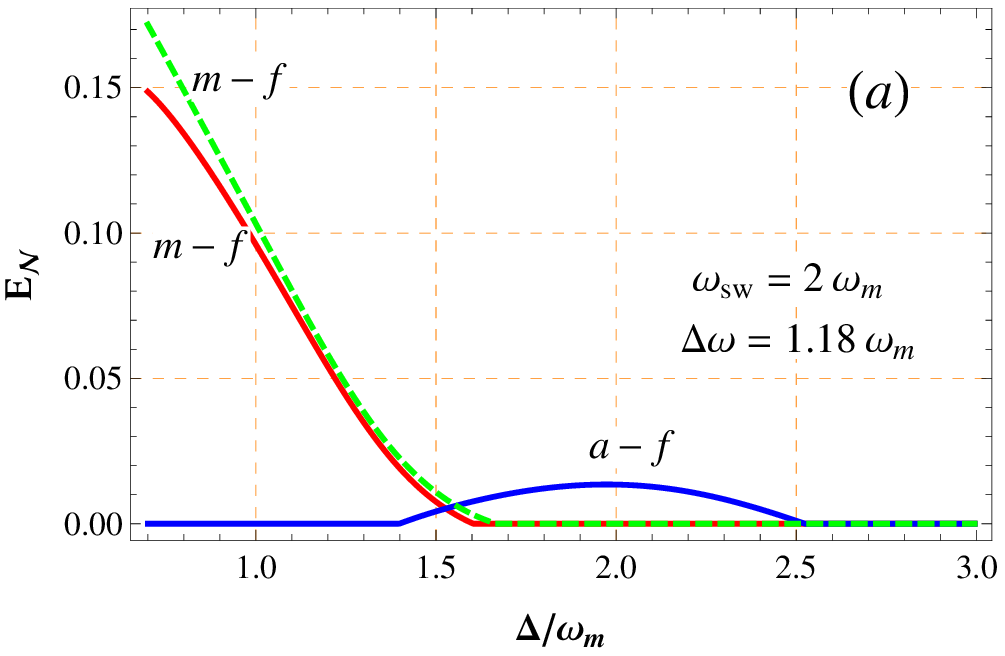}
\includegraphics[width=2.7in]{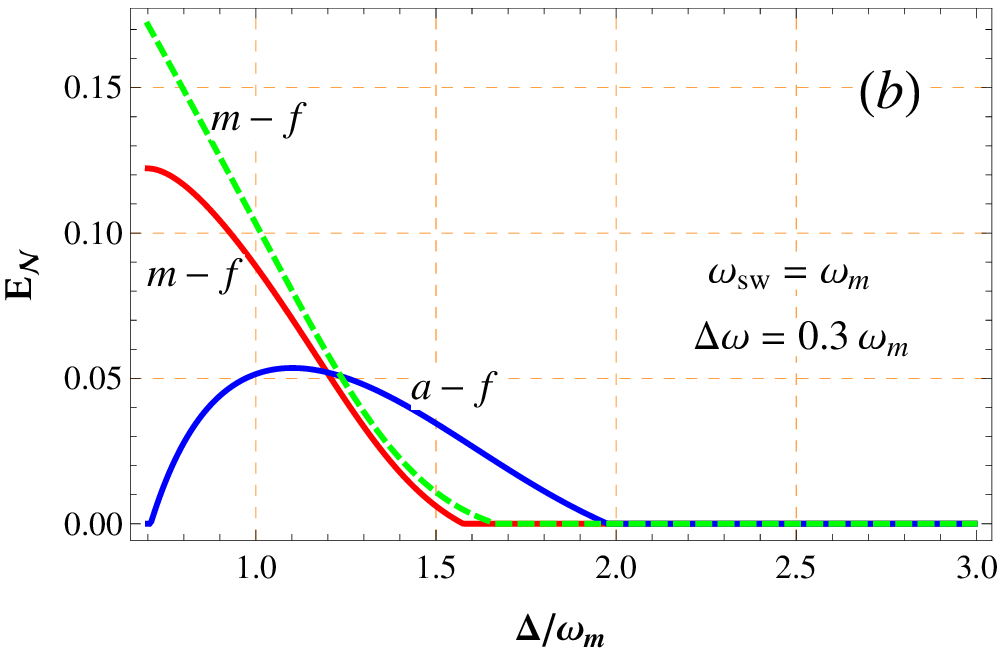}
\includegraphics[width=2.7in]{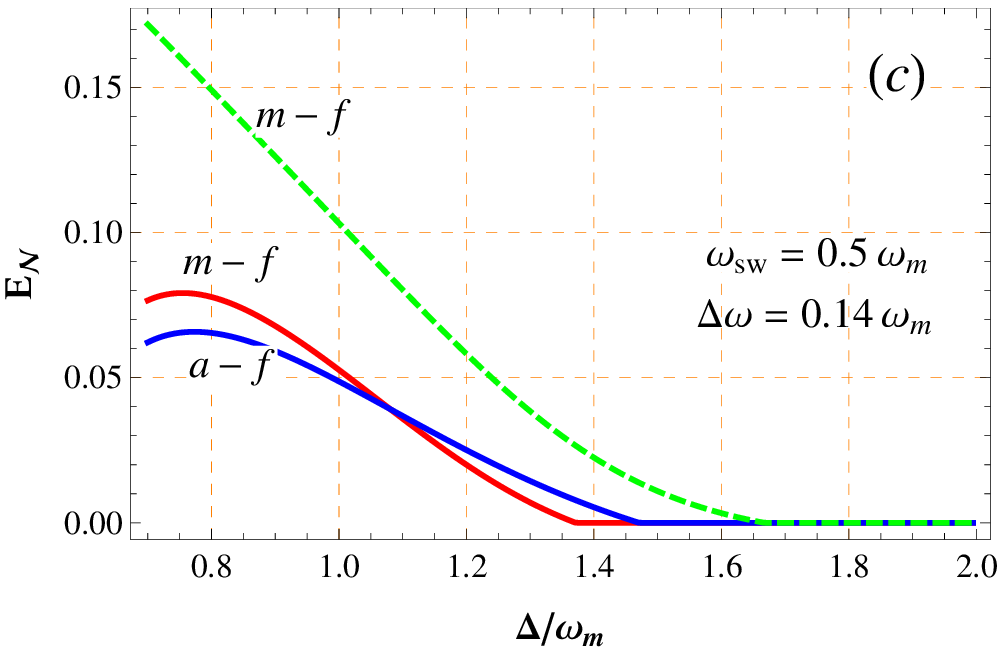}
\caption{
(Color online)
The bipartite mirror-field entanglement (\textit{m-f}) in the presence of BEC (red line) and in the absence of BEC (green dashed line) as well as the bipartite atom-field (\textit{a-f}) entanglement, i.e., the Bogoliubov mode and the field , (blue line) versus the normalized effective detuning $\Delta/\omega_{m}$ for three different values of the atom-atom interaction strength: (a) $\omega_{sw}=2\omega_{m}$, (b) $\omega_{sw}=\omega_{m}$ and (c) $\omega_{sw}=0.5\omega_{m}$ which leads to three different frequency differences $\Delta\omega$ between the mechanical and the Bogoliubov modes. All parameters are the same as those of Fig.\ref{fig:fig5}.}
\label{fig:fig6}
\end{figure}

In Fig.\ref{fig:fig5} we have plotted the effective incoherent number of mirror phonons as well as the effective incoherent number of atoms in the Bogoliubov mode versus the normalized detuning $\Delta/\omega_{m}$ for three different values of the atom-atom interaction strength. Furthermore, the bipartite mirror-field and atom-field entanglements have been represented in Fig.\ref{fig:fig6}.

The results obtained here are comparable to those of Ref.\cite{Genes2008} where the authors have investigated the problem of simultaneous cooling and entanglement of two mechanical modes of a mirror in an optical cavity. They have shown that cooling of the main mode depends on the frequency difference between the two modes. When this difference is equal or larger than the main mode frequency, the secondary mode does not disturb the cooling of the main mode.

Here, we study a similar system with the difference that we have considered the only vibrational mode of the mirror with resonance frequency $\omega_{m}$ as the main mode and the Bogoliubov mode of the BEC with resonance frequency $\omega_{B}=\sqrt{\Omega_{c}(\Omega_{c}+\omega_{sw})}$  as the secondary mode. Because of the dependence of the Bogoliubov mode frequency on $\omega_{sw}$, the present setup has the advantage that the resonance frequency of the secondary mode is controllable by the transverse frequency of the BEC trap.

Based on the dyanamical equation of the system, i.e., Eq.(\ref{nA}), the two so-called mechanical modes of the system with damping rates $\gamma_{m}=2\times10^{-5}\kappa$ and  $\gamma_{c}=10^{-3}\kappa$  are driven by the radiation pressure force with the effective frequency $\Delta$ and damping rate $\kappa=0.5 \omega_{m}$. It is well-known that if the frequency difference of the two mechanical modes $\Delta\omega=\omega_{B}-\omega_{m}$ is larger than the band width of the driven force $\kappa$, then the two modes can be cooled down to their ground states at two distinct ranges of effective detuning $\omega_{j}-\kappa<\Delta<\omega_{j}+\kappa,$ where the index $j$ refers to $m$ and $B$ for the mechanical and Bogoliubov modes, respectively \cite{Genes2008,Rae,Marquardt,Danata}. On the other hand, if $\Delta\omega<\kappa$, then both modes can be excited simultaneously by the radiation pressure at a fixed effective detuning $\Delta$. In this situation since the energy of the radiation pressure force is shared between the two modes simultaneously, the amount of energy exchanged between each mechanical mode and the optical field is reduced in comparison to the previous case. In this way the two cooling processes interfere destructively and neither the main nor the secondary mode is cooled effectively. If, in opposite, $\Delta\omega>\kappa$ the presence of BEC as the secondary mode does not affect the cooling of the main mode.

The situation $\Delta\omega>\kappa$ has been shown in Fig.\ref{fig:fig5} (a) where the frequency difference between two mechanical modes has been fixed to the value 1.18$\omega_{m}$ by fixing $\omega_{sw}=2\omega_{m}$.  As is seen, when $\Delta$ is near the mechanical frequency $\omega_{m}$, the effective number of mirror phonons in the presence of BEC (red line) goes to a minimum and the mirror is cooled down to its ground state without receiving any disturbance from the bogoliubov mode (note the complete overlap with the green dashed line which represents the effective incoherent mirror phonon number in the absence of BEC). For $\Delta\approx\omega_{B}\approx2.2\omega_{m}$, the incoherent number of atoms in the Bogoliubiv mode (blue line) is minimized which is somehow similar to a cooling process for the Bogoliubov mode as a mechanical oscillator.


On the other hand, the entanglement properties of the system, like the cooling process, depend on the frequency difference of the two mechanical modes. When $\Delta\omega$ is larger than the damping rate $\kappa$ of the cavity the presence of the secondary mode (the Bogoliubov mode) does not affect the entanglement of the main mode (mechanical mode) and the optical field. This situation has been shown in Fig.\ref{fig:fig6} (a) where $\omega_{sw}=2\omega_{m}$ and $\Delta\omega=1.18\omega_{m}$ (similar to Fig.\ref{fig:fig5} (a)). Here, again when $0.5\omega_{m}<\Delta<1.5\omega_{m}$, there is a complete overlap between the mirror-field entanglement in the presence of BEC (red line) and in the absence of BEC (green dashed line). Besides, when $\Delta$ gets near to the neighbourhood of $\omega_{B}$ where $\delta n_{c}$ is minimized, the mirror-field entanglement disappears and instead, the atom-field entanglement appears (blue line).

In Figs.\ref{fig:fig5}(b) and \ref{fig:fig6}(b), where $\omega_{sw}$ has been fixed at $\omega_{m}$, the frequency difference between the two modes is reduced to $\Delta\omega=0.3\omega_{m}$ and therefore $\Delta\omega<\kappa$. As is seen from Fig.\ref{fig:fig5} (b) the minimum of $\delta n_{c}$ (blue line) gets near to that of $\delta n_{m}$ (red line) and so the two modes can be excited and cooled simultaneously as $\Delta \rightarrow \omega_{m}$ . Besides, the presence of BEC has caused the incoherent number of mirror phonons (red line) to be a little different from that in the absence of BEC (green dashed line). Similarly, the mirror-field entanglement in the presence of BEC [red line in Fig.\ref{fig:fig6} (b)] has more deviation from that in the absence of BEC (green dashed line) and has also been diminished compared to its correspondence in Fig.\ref{fig:fig6} (a) while the atom-field entanglement has a noticeable increase.

In Figs.\ref{fig:fig5}(c) and \ref{fig:fig6}(c) the frequency difference between the two modes has been reduced to $0.14\omega_{m}$ by reducing $\omega_{sw}$ to $0.5\omega_{m}$. In this way the resonance frequencies of the two modes gets nearer to each other and therefore the presence of BEC affects the cooling and entanglement processes more strongly.  In Fig.\ref{fig:fig5} (c), $\delta n_{m}$ in the presence of BEC (red line) is completely different from that in the absence of BEC (green dashed line). Here, since the radiation pressure force is exciting the two modes with very close resonance frequencies simultaneously, neither of them could be cooled effectively.  Fig.\ref{fig:fig6} (c) shows that the mirror-field entanglement in the presence of BEC (red line) reduces and instead the atom-field entanglement (blue line) increases and gets near to the red line. Therefore, for small values of $\omega_{sw}$ the frequency difference $\Delta\omega$ reduces and the two bipartite entanglements get nearer to each other. This is similar to the result obtained in Ref.\cite{Paternostro} where the mirror-field entanglement can be measured from the the atom-field entanglement. Besides, in the range of parameters considered here the mirror-atom entanglement is zero.

\begin{figure}[ht]
\centering
\includegraphics[width=2.7in]{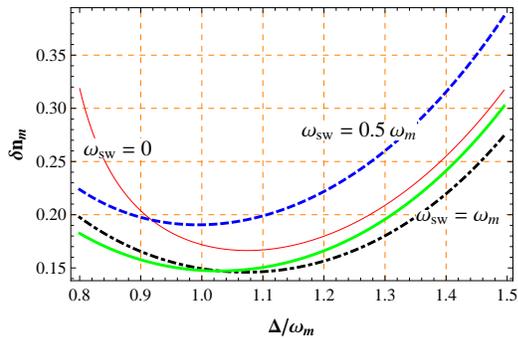}
\caption{
(Color online)
The incoherent number of mirror phonons versus the normalized detuning $\Delta/\omega_{m}$ for three different values of the atom-atom interaction strength: $\omega_{sw}=0$ (red thin line), $\omega_{sw}=0.5\omega_{m}$ (blue dashed line) and $\omega_{sw}=\omega_{m}$ (black dashed-dotted line). The green thick line corresponds to the absence of BEC.}
\label{fig:fig7}
\end{figure}

The important result obtained from Figs.\ref{fig:fig5} and \ref{fig:fig6} is the derivation of an interpretation for the concept of cooling of the Bogoliubov mode: the Bogoliubov mode is cooled when the effective incoherent number of atoms $\delta n_{c}$ is minimized (just like what happens for the mirror when $\delta n_{m}$ is minimized). In fact when $\Delta$ gets near to the resonance frequency of the Bogoliubov mode, $\delta n_{c}$ is minimized and simultaneously the atom-field entanglement is maximized. Therefore, the Bogoliubov mode behaves just like a secondary mechanical mode with the difference that the atoms play the role of phonons.

In order to see the effect of the \textit{s}-wave scattering interaction on the cooling of the mirror more precisely, in  Fig.\ref{fig:fig7} we have plotted the incoherent number of mirror phonons versus $\Delta/\omega_{m}$ for three different \textit{s}-wave scattering frequencies: $\omega_{sw}=0$ (red thin line), $\omega_{sw}=0.5\omega_{m}$ (blue dashed line), and $\omega_{sw}=\omega_{m}$ (black dashed-dotted line). The green thick line  represents the mirror phonon number in the absence of BEC. As is seen from Fig.\ref{fig:fig7} the presence of BEC does not help the cooling process of the mirror very much and even increases the mirror phonons when $\omega_{sw}$ is very small. The best situation occurs for $\omega_{sw}=\omega_{m}$ where the minimum of $\delta n_{m}$ in the presence and in the absence of BEC gets near to each other. In this case for $\Delta>\omega_{m}$ the cooling process is better than that in the absence of BEC. Similar results have very recently been obtained using the Bose-Hubbard model for simulating the BEC  by expanding the atomic wave field in terms of the Wannier functions \cite{Bhaarixiv}.

Based on the results obtained in  Fig.\ref{fig:fig6} and our previous explanations, the \textit{s}-wave scattering interaction can be used as a control parameter to change the frequency difference $\Delta\omega$ between the two modes. The larger $\Delta\omega$ the less the effect of BEC on the cooling of the mirror and the mirror-field entanglement. As is seen from Fig.\ref{fig:fig6} the mirror-field entanglement is increased by increasing the \textit{s}-wave scattering frequency. For $\omega_{sw}=2\omega_{m}$ it overlaps with the mirror-field entanglement in the absence of BEC. Inversely, the atom-field entanglement decreases by increasing the \textit{s}-wave scattering interaction and its maximum is shifted toward the larger values of $\Delta$.

\section{Conclusions}
In conclusion, we have studied a system consisting of a one-dimensional Bose-Einstein condensate inside a driven optomechanical cavity with a moving end mirror. In the weakly interacting regime, where just the first two symmetric momentum side modes with momenta $\pm2\hbar k$ are excited by fluctuations resulting from the atom-light interaction, the BEC can be considered as a one mode oscillator in the Bogoliubov approximation. In this way the Bogoliubov mode of the BEC is coupled to the optical field through a radiation pressure term which behaves as a secondary mechanical mode relative to the vibrational mode of the mirror (main mode).

We have investigated the effect of \textit{s}-wave scattering atom-atom interaction as well as the optomechanical coupling of the mirror on the optical bistablity of the cavity and on the cooling and entanglement processes of the two so-called mechanical modes. 

Firstly, we have shown that increasing \textit{s}-wave scattering interaction leads to a shift of the bistability region to the higher values along the parallel pump axis while the optomechanical coupling parameter of the mirror behaves inversely. In this way, one can control the bistability of the system by the \textit{s}-wave scattering frequency $\omega_{sw}$ which itself is controllable by the transverse trapping frequency $\omega_{\mathrm{\perp}}$. Therefore, it can be considered as another possibility of an optical switch realization in an optomechanical cavity.

Secondly, we have studied the effect of the presence of BEC and the \textit{s}-wave scattering interaction on the cooling of the mirror and the bipartite mirror-field and atom-field entanglements. It has been shown that if the frequency difference between the two mechanical modes $\Delta\omega$ is larger than the band width of the cavity $\kappa$, then the two modes can be cooled down to their ground states at two distinct ranges of effective detuning $\Delta$ and the presence of BEC does not affect the cooling of the mirror. Based on our results, when $\Delta$ gets near to the frequency of the Bogoliubov mode the effective incoherent numbers of atoms $\delta n_{c}$ is minimized and the atom-field entanglement is maximized. So, minimizing the effective incoherent numbers of atoms can be interpreted as the cooling of the Bogoliubov mode. On the other hand when $\Delta\omega<\kappa$ neither the main mode (mirror) nor the secondary mode (Bogoliubov mode) is cooled effectively. This setup has the advantage of controlling $\Delta\omega$ through $\omega_{sw}$ which is controllable by the transverse trapping frequency $\omega_{\mathrm{\perp}}$.

\section*{Acknowledgement}
The authors wish to thank The Office of Graduate
Studies of The University of Isfahan for their support.

\bibliographystyle{apsrev4-1}

\begin{thebibliography}{10}

\expandafter\ifx\csname natexlab\endcsname\relax\def\natexlab#1{#1}\fi
\expandafter\ifx\csname bibnamefont\endcsname\relax
  \def\bibnamefont#1{#1}\fi
\expandafter\ifx\csname bibfnamefont\endcsname\relax
  \def\bibfnamefont#1{#1}\fi
\expandafter\ifx\csname citenamefont\endcsname\relax
  \def\citenamefont#1{#1}\fi
\expandafter\ifx\csname url\endcsname\relax
  \def\url#1{\texttt{#1}}\fi
\expandafter\ifx\csname urlprefix\endcsname\relax\def\urlprefix{URL }\fi
 

\bibitem{LaHaye} M. D. LaHaye, O. Buu, B. Camarota, and K. C. Schwab, Science \textbf{304}, 74 (2004).
\bibitem{Abramovici} A. Abramovici \textit{et al.}, Science \textbf{256}, 325 (1992); P. Fritschel, Proc. SPIE \textbf{4856}, 282 (2003).
\bibitem{marshall} W. Marshall, C. Simon, R. Penrose, and D. Bouwmeester, phys. Rev. Lett. \textbf{91}, 130401 (2003); K. C. Schwab and M. L. Roukes, Phys. Today \textbf{58} (7), 36 (2005).
\bibitem{Genes 2008} C. Genes, D. Vitali, P. Tombesi, S. Gigan, and M. Aspelmeyer, Phys. Rev. A \textbf{77}, 033804 (2008).
\bibitem{Teufel} J. D. Teufel, J. W. Harlow, C. A. Regal, and K. W. Lehnert, Phys. Rev. Lett. \textbf{101}, 197203 (2008); J. D. Teufel \textit{et al.}, Nature (London) \textbf{475}, 359 (2011); J. D. Teufel \textit{et al.}, Nature (London) \textbf{471}, 204 (2011).
\bibitem{Liberato} S. De Liberato, N. Lambert, and F. Nori, Phys. Rev. A \textbf{83}, 033809 (2011).
\bibitem{Barzanjeh2} Sh. Barzanjeh, M. H. Naderi, and M. Soltanolkotabi Phys. Rev. A \textbf{84}, 023803 (2011).
\bibitem{Brenn Nature} F. Brennecke, T. Donner, S. Ritter, T. Bourdel, M. Kohl, and T. Esslinger, Nature (London) \textbf{450},268 (2008).
\bibitem{Gupta} S. Gupta, K.L. Moore, K.W. Murch, and D.M. Stamper-Kurn, Phys. Rev. Lett. \textbf{99}, 213601 (2007).
\bibitem{Brenn Science} F. Brennecke, S. Ritter, T. Donner, and T. Esslinger, Science \textbf{322}, 235 (2008).
\bibitem{Masch Ritch 2004} C. Maschler and H. Ritsch, Opt. Commun. \textbf{243}, 145 (2004).
\bibitem{Dom JB}P. Domokos, P. Horak, and H. Ritsch, J. Phys. B, At. Mol. Opt. Phys. \textbf{34}, 187, (2001).
\bibitem{Kanamoto 2010} R. Kanamoto and P. Meystre, Phys. Scr. \textbf{82}, 038111 (2010).
\bibitem{Nagy Ritsch 2009} D. Nagy, P. Domokos, A. Vukics, and H. Ritsch, Eur. Phys. J. D \textbf{55}, 659 (2009).
\bibitem{Genes2008}C. Genes, D. Vitali, and P. Tombesi, New J. Phys. \textbf{10}, 095009(2008).
\bibitem{Morsch} O. Morsch and M. Oberthaler, Rev. Mod. Phys. \textbf{78}, 179 (2006).
\bibitem{meys}P. Meystre, E. M. Wright, J. D. McCullen, and E. Vignes, J. Opt. Soc. Am. B\textbf{2}, 1830 (1985).
\bibitem{Szirmai 2010}G. Szirmai, D. Nagy, and P. Domokos, Phys. Rev. A \textbf{81}, 043639 (2010).
\bibitem{Ritter Appl. Phys. B} S. Ritter, F. Brennecke, K. Baumann, T. Donner, C. Guerlin, and T. Esslinger, Appl. Phys. B \textbf{95}, 213 (2009).
\bibitem{Zubairy}S. Yang, M. Al-Amri, J. Evers, and M. S. Zubairy, Phys. Rev. A \textbf{83}, 053821 (2011).
\bibitem{new nagy} D. Nagya, G. Szirmai, and P. Domokos, e-print arXiv:1303.2977v1 (2013).
\bibitem{Steinke Meustre 2011} S. K. Steinke and P. Meystre, Phys. Rev. A \textbf{84}, 023834 (2011).
\bibitem{Chen Meystre 2010} W. Chen, D. S. Goldbaum, M. Bhattacharya, and P. Meystre, Phys. Rev. A \textbf{81}, 053833 (2010).
\bibitem{Konya Domokos 2011} G. Konya, G. Szirmai, and P. Domokos, Eur. Phys. J. D \textbf{65}, 33 (2011).
\bibitem{Gardiner}C. W. Gardiner and P. Zoller, \textit{Quantum Noise} (Springer, Berlin, 2000).
\bibitem{Gigan} S. Gigan, \textit{et al.}, Nature (London) \textbf{444}, 67 (2006).
\bibitem{Arcizet}O. Arcizet, P.-F. Cohadon, T. Briant, M. Pinard, and A. Heidmann, Nature (London) \textbf{444}, 71 (2006).
\bibitem{Rgenes}C. Genes, D. Vitali, and P. Tombesi, Phys. Rev. A \textbf{77}, 050307(R) (2008).
\bibitem{Dalafi}A. Dalafi, M. H. Naderi, M. Soltanolkotabi, and Sh. Barzanjeh, Phys. Rev. A \textbf{87}, 013417 (2013).
\bibitem{Zhang 2009} J. M. Zhang, F. C. Cui, D. L. Zhou, and W. M. Liu, Phys. Rev. A \textbf{79}, 033401 (2009).
\bibitem{Horak 2000} P. Horak, S. M. Barnett, and H. Ritsch, Phys. Rev. A \textbf{61}, 033609 (2000).
\bibitem {RH} I. S. Gradshteyn and I. M. Ryzhik, \textit{Table of Integrals, Series
and Products} (Academic Press, Orlando, 1980); A. Hurwitz, \textit{Selected Papers on Mathematical Trends in Control Theory}, edited by R. Bellman and R. Kabala (Dover, New York, 1964).
\bibitem{eis}J. Eisert, Ph.D. thesis, University of Potsdam, 2001; G. Vidal and R. F. Werner, Phys. Rev. A \textbf{65}, 032314 (2002); G. Adesso, A. Serafini, and F. Illuminati, Phys. Rev. A \textbf{70}, 022318 (2004).
\bibitem{Rae}I. Wilson-Rae, N. Nooshi, W. Zwerger, and T. J. Kippenberg, Phys. Rev. Lett. \textbf{99}, 093901 (2007).
\bibitem{Marquardt}F. Marquardt \textit{et al.}, Phys. Rev. Lett. \textbf{99}, 093902 (2007).
\bibitem{Danata}A. Dantan \textit{et al.}, Phys. Rev. A \textbf{77}, 011804 (2008).	
\bibitem{Paternostro}G. De Chiara, M. Paternostro, and G.M. Palma, Phys. Rev. A \textbf{83}, 052324 (2011).
\bibitem{Bhaarixiv}S. Mahajan, N. Aggarwal, A. Bhattacherjee, and ManMohan, e-print arXiv:1302.0339v1 (2013).

\end{thebibliography}

\end{document}